\documentclass[useAMS,usenatbib]{mn2e}
\usepackage{url}
\usepackage{graphicx}
\usepackage{verbatim}
\usepackage[usenames]{color}
\DeclareGraphicsExtensions{.pdf,.png,.jpg,.mps,.eps,.ps}
\usepackage{amsmath}
\usepackage{natbib}
\usepackage{color}
\usepackage{graphicx}
\usepackage{lscape}

\newcommand\nustar{{\it NuSTAR}}
\newcommand\swift{{\it Swift}}

\newcommand\suzaku{{\it Suzaku}}
\newcommand\rxte{{\it RXTE}}
\newcommand\xmm{{\it XMM-Newton}}
\newcommand\inte{{\it INTEGRAL}}

\newcommand\kev{{\rm~keV}}

\newcommand\kms{\ifmmode {\rm~km\ s}^{-1} \else ~km s$^{-1}$\fi}
\newcommand\Hunit{\ifmmode {\rm~km\ s}^{-1}\ {\rm Mpc}^{-1}
        \else ~km s$^{-1}$ Mpc$^{-1}$\fi}
\newcommand\ctssec{\ifmmode {\rm~count\ s}^{-1} \else ~count s$^{-1}$\fi}
\newcommand\ergsec{\ifmmode {\rm~erg\ s}^{-1} \else
        ~erg s$^{-1}$\fi}
\newcommand\funit{\ifmmode {\rm~erg\ s}^{-1}\;{\rm cm}^{-2} \else
        ~ergs s$^{-1}$ cm$^{-2}$\fi}
\newcommand\phflux{\ifmmode {\rm~photon\ s}^{-1}\;{\rm cm}^{-2}
        \else   ~photon s$^{-1}$ cm$^{-2}$\fi}
\newcommand\efluxA{\ifmmode {\rm~erg\ s}^{-1}\;{\rm cm}^{-2}\;{\rm
        \AA}^{-1} \else ~erg s$^{-1}$ cm$^{-2}$ \AA$^{-1}$\fi}
\newcommand\efluxHz{\ifmmode {\rm~erg\ s}^{-1}\;{\rm cm}^{-2}\;{\rm
        Hz}^{-1} \else ~erg s$^{-1}$ cm$^{-2}$ Hz$^{-1}$\fi}
\newcommand\cc{\ifmmode {\rm~cm}^{-3} \else cm$^{-3}$\fi}
\newcommand\FWHM{\ifmmode {\rm~FWHM} \else ${\rm~FWHM}$\fi}
\newcommand\Msun{\ifmmode M_{\odot} \else $M_{\odot}$\fi}
\newcommand\Lsun{\ifmmode L_{\odot} \else $L_{\odot}$\fi}
\newcommand\ltsim{\raisebox{-.5ex}{$\;\stackrel{<}{\sim}\;$}}
\newcommand\gtsim{\raisebox{-.5ex}{$\;\stackrel{>}{\sim}\;$}}
\newcommand\hbeta{\ifmmode {\rm H}\beta \else H$\beta$\fi}
\newcommand\Kalpha{\ifmmode {\rm K}\alpha \else K$\alpha$\fi}
\newcommand\nh{\ifmmode N_{\rm H} \else N$_{\rm H}$\fi}

\usepackage{graphicx}
\usepackage{url}
\usepackage{verbatim}
\usepackage{color}

\title[\suzaku{} and \nustar{} spectroscopy of 4U~1820--30]{Broadband X-ray emission and the reality of the broad iron line from the Neutron Star -- White Dwarf X-ray binary 4U~1820--30}
\author[Mondal et al.] {\parbox[]{6.5in}{{Aditya S. Mondal$^{1}\thanks{E-mail: adityas.mondal@visva-bharati.ac.in}$, G. C. Dewangan$^{2}$,
  M. Pahari$^{2}$, R. Misra$^{2}$, A. K. Kembhavi$^{2}$,  B. Raychaudhuri$^{1}$} \\ 
  \small
$^{1}$Department of physics, Visva-Bharati, Santiniketan, West Bengal-731235, India \\
$^{2}$Inter-University Centre for  Astronomy \& Astrophysics (IUCAA), Pune, 411007 India
}}

\date{\today}
\begin{document}
\maketitle
\begin{abstract}
Broad relativistic iron lines from neutron star X-ray binaries are important probes of the inner accretion disk. The X-ray reflection
features can be weakened due to strong magnetic fields or very low iron abundances such as is possible in X-ray binaries with low mass,
first generation stars as companions. Here we investigate the reality of the broad iron line detected earlier from the neutron star low mass X-ray binary 4U~1820--30 with a degenerate helium dwarf companion.  We perform a comprehensive, systematic broadband spectral study of the atoll source using \suzaku{} and  simultaneous  \nustar{} \& \swift{} observations.  We have used different continuum models involving
accretion disk emission, thermal blackbody and thermal Comptonization of either disk or blackbody photons. The \suzaku{} data show positive and negative residuals in the region of iron K band. These features are well described by two absorption edges at $7.67\pm0.14\kev$ and $6.93\pm0.07\kev$ or partial covering photoionized absorption  or by blurred reflection. Though, the simultaneous \swift{} and \nustar{} data do not clearly reveal the emission or absorption features, the data are consistent with the presence of either absorption or emission features. Thus, the absorption based models provide an alternative to the broad iron line or reflection model. The absorption features may arise in winds from the inner accretion disk. The broadband spectra appear to disfavour continuum models in which the blackbody emission from the neutron star surface provides the seed photons for thermal Comptonization. Our results suggest emission from a thin accretion disk ($kT_{disk} \sim 1\kev$), Comptonization of disk photons in a boundary layer most likely covering a large fraction of the neutron star surface and innermost parts of the accretion disk, and blackbody emission ($kT_{bb} \sim 2\kev$) from the polar regions.

\end{abstract}
\section{introduction}

A low mass X-ray binary (LMXB) system consists of a low mass companion star ($\ltsim 1 M_{\odot}$) and a primary compact star (a neutron star or a black hole) rotating around each other. For such a system, in the course of evolution when the companion star fills its roche lobe, matter is transferred from the companion star to the compact star. This matter follows a slow spiral path and forms an accretion disc around the primary \citep{2002apa..book.....F}. For the supermassive as well as the stellar-mass black holes, X-ray emission lines specifically, Fe K$\alpha$ line $\sim6.4\kev$, from the inner part of the accretion disc are well-known \citep{2007ARA&A..45..441M}. With the advance in X-ray spectroscopy, asymmetry in the Fe K$\alpha$ line profile has been observed in many sources including persistent neutron star (NS) low mass X-ray binaries. Several authors have discussed the inner disc origin of such a line from the different NS LMXBs and attributed the line asymmetry to relativistic blurring \citep{2007ApJ...664L.103B, 2008ApJ...674..415C, 2008ApJ...688.1288P, 2009A&A...493L..39P, 2009MNRAS.399L...1R}. 

4U~1820--30 is an extensively studied source because of its extraordinary  properties. It is an ultracompact NS X-ray binary with an orbital period of 11.4 min \citep{1987ApJ...312L..17S}. Thermonuclear X-ray bursts from this source were first discovered by \citet{1976ApJ...205L.131G} and confirmed that the compact object as a neutron star which resides in the metal-rich globular cluster NGC~6624. Based on optical observation of this cluster, \citet{1985PASP...97..465H} estimated the distance to be $7.6$ kpc. The binary system has a low mass ($\sim 0.07 M_{\odot}$), roche-lobe filling, degenerate Helium ($He$) dwarf as a companion and the accreted matter likely has a very high $He$ abundances \citep{1987ApJ...322..842R}. Based on theoretical modelling, \citet{2003ApJ...595.1077C} showed that the bursts from 4U~1820--30 are consistent with a picture of accumulation and burning of $He$-rich material. The mass of the NS in this binary is not very well known. \citet{2010ApJ...719.1807G} estimated the mass of the NS to be $\sim 1.6 M_{\odot}$ based on the spectral analysis of multiple thermonuclear X-ray bursts. However, \citet{2013MNRAS.429.3266G} found that the decaying phase of those thermonuclear bursts did not satisfy the relation $F\propto T_{bb}^4$ which was a crucial assumption to estimate the mass of the NS. \citet{1997ApJ...482L..69A} reported the system inclination angle $i$ to be in the range of $35^0\le i \le 50^0$.

Rossi X-ray timing explorer (\rxte{}) observation of the atoll source 4U~1820--30 have shown kilohertz (kHz) quasi-periodic oscillations (QPOs) at different frequencies in its power density spectrum (PDS). \citet{1998ApJ...500L.171Z} reported the highest QPO frequency to be $1060\pm20$ Hz from 4U~1820--30 which may be the signature of the marginally stable orbit. The source also exhibited a $176$ day periodic modulation in the X-ray flux \citep{1984ApJ...284L..17P}. Photospheric radius expansion (PRE) X-ray burst have been reported for this source from the analysis of \rxte{} data \citep{2008ApJS..179..360G}. On September 1999, a remarkable 3 hr superburst was observed from 4U~1820--30, during which its peak isotropic luminosity was $\sim 3.4 \times 10^{38}$ ergs/s assuming a distance of $6.6$ kpc. Spectral analysis of this superburst revealed the evidence of a broad emission line between $5.8$ and $6.4$ keV and an absorption edge at $7-9$ keV \citep{2002ApJ...566.1045S}.  \citet{2012A&A...539A..32C} fitted the broadband continuum  with a blackbody and a Comptonization model obtained from \xmm{} EPIC-PN and \inte{} data taken during April 2009. However, they found no clear evidence of iron emission line in the $6.4$ keV region.

\citet{2008ApJ...674..415C} performed the spectral analysis of the \suzaku{} data taken on 14 Sept. 2006 using the XIS in the $1-9$ keV band and the $12-25$ keV band of the HXD/PIN (ignoring $1.5-2.5$ keV region). By using a variety of continuum models, \citet{2008ApJ...674..415C} detected a significant iron line and fitted it to a relativistic iron line model by restricting the line energy to be between $6.4-6.97$ keV. \citet{2010ApJ...720..205C} reanalyzed the data with improved calibration after combining all the spectra obtained with the front illuminated CCDs (for higher signal to noise ratio) and reported significant (but weaker than before) detection of the emission line at $\sim 6.4 $ keV. While \citet{2008ApJ...674..415C} reported the presence of iron line profile at energies $\sim 6.4$ keV from the \suzaku{} spectral fitting, \citet{2012A&A...539A..32C} found no significant evidence of iron emission line at energies $\sim 6.4$ keV in the spectra obtained from combined \inte{} and \xmm{} data. As the source 4U~1820--30 is a neutron star and $He$ white dwarf binary, the presence of the iron K$\alpha$ emission line provides crucial information on the nature of the white dwarf and hence it is important to verify the reality of any iron line features in its X-ray spectra. \\

A disadvantage of the X-ray CCD detectors is the photon that pile-up can distort the shape of the spectra especially in the Fe K band. Thus, it is important to test the spectral properties using broadband spectra, that are either free of photon pile-up distortions or properly pile-up corrected. \nustar{} with its pile-up free operation in the $\sim 3-79$ keV provides such an opportunity. In this work, we re-analyse the spectra obtained from \suzaku{} and test the results using the new simultaneous \nustar{} and \swift{} data. The \suzaku{} data and the combined \nustar{} and \swift{} data provide broad band coverage from $0.3-30$ keV allowing us to constrain the different broad spectral components as well as an iron line feature. We organised the paper as follows. In Sec. 2 we describe the observation details and data reduction, in Sec.3 we discuss the temporal behaviour of the source and in Sec.4 we discuss the details of spectral modeling. Finally, we summarize our results and conclusions in Sec.5.

\section{observation and data reduction}
\subsection{\suzaku{} XIS and HXD/PIN}
The \suzaku{} observation of the source 4U~1820--30 was performed on 14th Sept. 2006 at the X-ray Imaging Spectrometer (XIS; \citealt{2007PASJ...59S..23K}) nominal pointing position with XIS detectors operating in $1/4$ window mode. The net exposure after instrumental deadtime were about $13$ ks and $29.6$ ks for XIS and PIN (positive intrinsic negative), respectively. The observation details are summarized in Table~\ref{obs_summary}. The data reduction was performed using the convenient and straightforward
analysis tool {\tt XSELECT V2.4c}. \\ 
As 4U~1820--30 is one of the brightest known X-ray sources, there may be some significant pile up (up to $20\%$ or so) in the XIS detectors. Therefore, we checked for the presence of a significant pile-up. We applied the attitude correction using the tool {\tt aeattcorr.sl} to reduce the pointing determination error for this bright source and used the attitude corrected event file for further analysis. We also ran another routine {\tt pile-estimate.sl} which allowed us to create a region file that excluded the most piled up areas and then to estimate the effective pile up fraction of the remaining events. We selected an annulus region of outer radius $140 \arcsec$ and inner radius $30 \arcsec$ and then gradually increased the inner radius. We found that when we excluded the region of $60\arcsec$ radius, the effective pile up fraction was $\sim 5 \%$. We extracted the source spectrum after selecting an annulus region of outer radius $140\arcsec$ and inner radius $60\arcsec$. Similarly, we extracted a background spectrum from a $127\arcsec$ source free circular region. We then created the response files accordingly for each XIS detector,  after which  we combined the spectra of all front-illuminated detectors (XIS0, XIS2 and XIS3) to get higher signal-to-noise ratio using the tool {\tt addascaspec}. Finally, the spectra were rebinned by a factor of $4$ to enhance the signal-to-noise ratio by a factor $\sim2$.   \\ 

In the case of Hard X-ray Detector (HXD; \citealt{2007PASJ...59S..35T}) PIN data, we downloaded the latest {\it tuned} background file from the \suzaku{} web site.\footnote{http://heasarc.gsfc.nasa.gov/docs/suzaku/analysis/pinbgd.html} We ran the ftool {\tt hxdpinxbpi} to produce the dead time corrected PIN source spectrum as well as the PIN background spectrum. \\

\subsection{\nustar{} FPMA and FPMB}
\nustar{} observed the source 4U~1820--30 on 2013 July 11 (Obs. ID: $80001011002$). \nustar{} data were collected with the focal plane module
telescopes (FPMA and FPMB). The net exposure after instrumental deadtime was about 1.8 ks. Table~\ref{obs_summary} lists the observation details. \\
 
The data reduction was performed using the \nustar{} data analysis software ({\tt NuSTARDAS v1.4.1}) with the latest calibration version $20141020$. We filtered the event file and applied the default depth correction using the {\tt nupipeline} task. The source region was
extracted from a $100\arcsec$ circular region centered on the source position. The background region was taken from the corner of the same
detector as close as possible to the source without including the source photons. Source and background light curves and spectra were extracted from the FPMA and FPMB data and response files were generated using the {\tt nuproduct} task. The data from the two FPM were modelled
simultaneously in order to minimize systematic effects. Initially, we rebinned the spectra by a  minimum of $100$ counts per bin but the data appeared to be noisy for such a rebinning and we finally rebinned after taking a minimum of $800$ counts per bin. 

\subsection{\swift{}/XRT}

The source was also observed by \swift{} nearly simultaneously with the \nustar{} observation, the details of which are provided in Table 1.
Following standard procedure for filtering and screening, \swift{} X-ray ray telescope (XRT) data were reduced using the {\tt xrtpipeline v 0.13.0} tool. The data obtained from photon counting (PC) mode were found to be severely piled up. Data extracted from a $\sim$ 30$\arcsec$ circular region in the image had an average count rate of 102.3 $\pm$ 5.4 counts/s. Therefore, the PC mode data was not useful. \swift{} also observed this source using Window Timing (WT) mode with an exposure of $\sim$1.9 ks. The background subtracted average count rate in this mode was found to be 81.7 $\pm$ 2.5 counts/s which is well below the prescribed photon pile-up limit for the WT mode data ($\sim$ 100 counts/s; \citealt{2006A&A...456..917R}). Therefore, source and background spectra were extracted using rectangular boxes with {\tt XSELECT v 2.4}. The latest \swift{}/XRT spectral redistribution matrices were used. The {\tt xrtmkarf} tool was used along with exposure map file to generate an auxiliary response file for the current observation.

\section{Hardness - intensity diagram}
\begin{figure}
\centering
\includegraphics[width=8.4cm,angle=0]{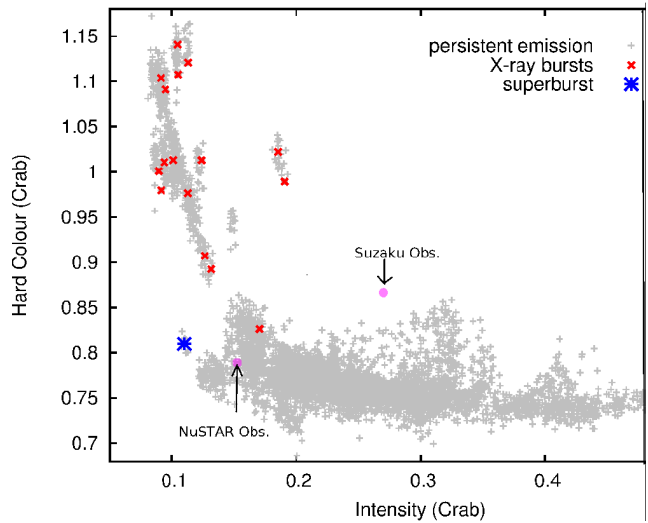}
\caption{Hardness - intensity diagram of all {\it RXTE} observations of 4U~1820--30 (\citealt{2013MNRAS.429.3266G}). Our \suzaku{} and \nustar{} observations are shown by arrows. The intensity and hard color are in crab units.}
  \label{fig_HID}
\end{figure}

 Following \citealt{2013MNRAS.429.3266G}, we generated hardness - intensity diagram by calculating the hard color as a flux ratio in the  $6 - 9.7\kev$ and $9.7 - 16\kev$ bands and the intensity as the $2 - 10\kev$  flux. 
We extracted the spectrum of the Crab from the \rxte{}/PCA data (Obs. ID: $92802-02-05-00$) taken close to our observations and calculated the fluxes in the above mentioned energy bands and used those to normalize the fluxes in each energy band for proper comparison. By such a procedure, we compared the spectral states during the \suzaku{} and the \nustar{} observations in the same hardness-intensity diagram as shown in Fig.~\ref{fig_HID}. For both the \suzaku{} and the \nustar{} observations the source was observed in the so-called banana branch (soft/high) spectral states of the atoll sources \citep{1989A&A...225...79H}. In the hardness - intensity diagram, we found that during \suzaku{} and \nustar{} observation the spectral changes occur mostly in flux and only little change in spectral hardness (see Fig.~\ref{fig_HID}).

\begin{table*}
  \caption{Observation details for the \suzaku{}, \nustar{} and \swift{} on 4U~1820--30} 
  
  \begin{tabular}{|p{2.3cm}|p{2.0cm}|p{2.3cm}|p{2.0cm}|p{2.0cm}|}
    \hline
    Instrument & Obs. ID. & Obs. start date (dd/mm/yyyy) & Exposure time(ks) & 
    Obs. Mode  \\ 
    \hline
    Suzaku/XIS & 401047010 & 14/09/2006 & 13.0 & 1/4 W \\
    Suzaku/PIN & 401047010 & 14/09/2006 & 29.6  & --  \\
    NuSTAR/FPMA, FPMB   & 80001011002 & 08/07/2013 & 1.8  & SCIENCE \\
	SWIFT/XRT & 00080210001 & 07/07/2013 & 1.9 & WT \\
    \hline 
  \end{tabular}\label{obs_summary} \\
\end{table*}

\begin{figure*}
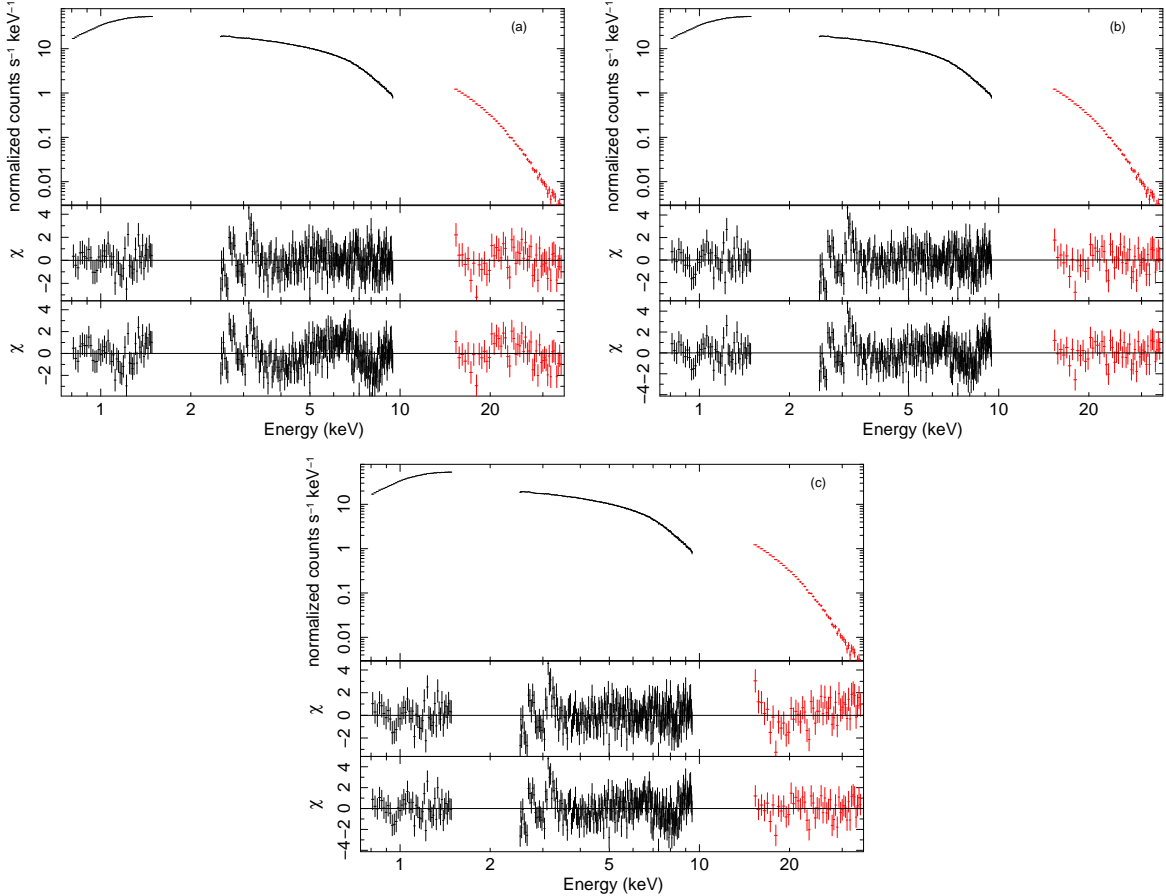

  \centering
  \includegraphics[scale=0.32,angle=-90]{fig2a.ps}
  \includegraphics[scale=0.32,angle=-90]{fig2b.ps}
  \includegraphics[scale=0.32,angle=-90]{fig2c.ps}
  \caption{{\it Suzaku} combined XIS and PIN spectral data with the
    best-fitting models (a) {\tt TBabs$\times$edge$\times$edge(diskbb + bbody + powerlaw)}, (b) {\tt TBabs$\times$edge$\times$edge
      (diskbb + bbody + nthcomp)} with the temperature ($kT_{seed}$) of the seed photons tied to the inner disc temperature ($kT_{disc}$) of the {\tt diskbb} component, and (c) same as (b) but $kT_{seed}$ tied to the temperature of {\tt bbody} $kT_{bb}$. Middle panels show the
    deviations from the best-fitting models. Lower panels show the deviations from the best-fitting models without the absorption edges.}
  \label{fig_suzaku_fit}
\end{figure*}

\section{spectral analysis}
\subsection{\suzaku{} spectra}
We analysed the spectra using  {\tt XSPEC} version {\tt $12.8.1g$} \citep{1996ASPC..101...17A}. We fitted the combined XIS(0, 2, 3) data in the $0.8-10$ keV band and the HXD/PIN data in the $15-35$ keV band simultaneously. We ignored the $1.5-2.5$ keV band due to  known calibration uncertainties. We used a constant factor to account for absolute flux calibration offset between the XIS and HXD/PIN. 

\subsubsection{Continuum Shape}

X-ray continuum of NS LMXBs can be equally well explained by different continuum models. Apart from a power-law or Comptonized component, the soft component can arise from the NS surface or the boundary layer. The soft component can also arise from the accretion disc around the NS \citep{2007ApJ...667.1073L}. We first tested a two component model; blackbody plus a diskbb ({\tt bbody} and {\tt diskbb} in {\tt XSPEC}, respectively, \citealt{1984PASJ...36..741M}). We accounted for the absorption by multiplying the absorption model {\tt TBabs} with abundance set to {\tt wilm} \citep{2000ApJ...542..914W}.  We also used a constant factor to account for any differences in the cross-calibration normalization between different instruments. We fixed this constant to 1 for the combined XIS and varied for HXD/PIN. The combination of these two models did not yield a good fit to the data ($\chi^2/dof=1116/333$ where $dof$ stands for degrees of freedom). Hence, we used the following continuum models which we describe below in details.  
\begin{enumerate}
\item Model 1: {\tt TBabs$\times$(diskbb $+$ bbody $+$ powerlaw)}   
\item Model 2: {\tt TBabs$\times$(diskbb $+$ bbody $+$ cuttofpl)}
\item Model 3: {\tt TBabs$\times$(diskbb $+$ bbody $+$ nthcomp)}
\item Model 4: {\tt TBabs$\times$(diskbb $+$ bbody $+$ CompTT)}
\end{enumerate} 

Model 1 provided an improved fit ($\Delta \chi^2 = -537$ for the addition of $2$ parameters) compared to the two component model {\tt TBabs$\times$(diskbb $+$ bbody)}. This implied that the {\tt power-law} played a significant role in fitting the continuum but the fit was formally not acceptable ($\chi^2/dof=579/331$). The best-fitting {\tt power-law} slope was $\Gamma=2.6\pm0.05$. The {\tt diskbb} and the {\tt bbody} temperature obtained were $kT_{disc}=1.27\pm 0.02$ keV  and $kT_{bb}=2.53\pm0.02$ keV, respectively. In the residuals we found an absorption feature at $\sim 7.5$ keV and an emission feature at $\sim 6.4$ keV.
The spectral data with residuals are shown in the top and lower panel of Fig.~\ref{fig_suzaku_fit}(a), respectively. 
 \\

\begin{figure}
\centering
\includegraphics[scale=0.35,angle=-90]{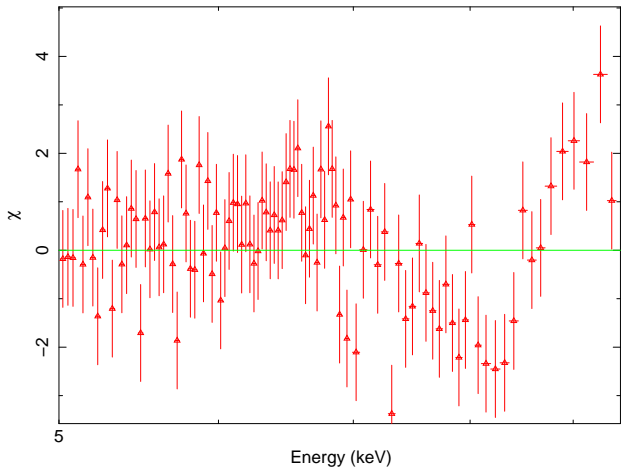}
\caption{ Absorption like features is present in the energy band $7 - 9\kev$ in the spectrum of 4U~1820--30 when the continuum is fitted with model 2, namely, {\tt TBabs$\times$(diskbb + bbody + cutoffpl)}.}
  \label{fig_edge}
\end{figure}

Following \citet{2000ApJ...542.1000B}, we tried to fit the continuum using a power-law with exponential cut-off ({\tt cutoffpl} in {\tt XSPEC}, an analytical approximation of unsaturated Comptonization) model along with the two thermal components {\tt diskbb} and {\tt bbody} i.e., model 2. This model {\tt TBabs$\times$(bbody + diskbb + cutoffpl)} fitted the continuum well ($\chi^2/dof=475/331$), but resulted in residuals in the $6 - 9\kev$ band which is shown in Fig.~\ref{fig_edge}. \\

Next we used a physical thermal Comptonization model {\tt nthcomp} \citep{1996MNRAS.283..193Z, 1999MNRAS.309..561Z}. The physical picture is that the hot electrons Compton upscatter seed photons with a(quasi)blackbody spectrum e. g., from a neutron star surface or an accretion disc. Either of these shapes can be selected via the model
parameter {\it input type} with both being parameterized by a seed photon temperature (denoted by $kT_{seed}$).  From the model 1, we replaced the {\tt power-law} component by the Comptonized model {\tt nthcomp} (model 3). 
We  first tied the temperature of the {\tt nthcomp} model ($kT_{seed}$) to the temperature of the {\tt diskbb} ($kT_{disc}$) model (defined as model 3A). This model provided $\chi^2$/dof=$473/330$.  We performed the fit again after tying the {\tt nthcomp} $kT_{seed}$ to the single temperature {\tt blackbody} $kT_{bb}$ (defined as model 3B). This model resulted in $\chi^2$/dof=$479.6/330$. Both the above choices for the seed photon temperature in the {\tt nthcomp} model resulted in an absorption edge like signature $\sim 7.5$ keV and an emission line feature  at $\sim6.4 $ keV (see Fig.~\ref{fig_suzaku_fit}(b) and (c), lower panel). \\

We also examined the continuum with another thermal Comptonization model {\tt CompTT} \citep{1994ApJ...434..570T} along with the same thermal components. The model 4, {\tt TBabs$\times$(bbody + diskbb + CompTT)} with $kT_{seed}$=$kT_{disc}$ provided us a consistent fit ($\chi^2$/dof=$472.6/330$) with what we obtained earlier with other continuum models. \\

Apart from the main continuum models discussed above, we tried some additional models to fit the continuum. We tested the combination of one thermal component and one Comptonized component.  We combined either {\tt diskbb} or {\tt bbody} to the {\tt nthcomp} component. We defined the combination of the {\tt diskbb} and the {\tt nthcomp} as model 0A and the the combination of the {\tt bbody} and the {\tt nthcomp} as model 0B. 
\begin{enumerate}
\item Model 0A: {\tt TBabs$\times$(diskbb + nthcomp)}
\item Model 0B: {\tt TBabs$\times$(bbody + nthcomp)}
\end{enumerate}
It may be noted that both the models provided us a significantly worse fit ($\chi^2$/dof=$500/332$ and $496/332$ for the model 0A and 0B, respectively) compared to the two thermal component model.  \\

\subsubsection{Modelling of the features seen $\sim 6-8\kev$ region} 
 
When we fitted the continuum with different combinations of phenomenological and physical models, in each case the residuals showed an absorption edge at $\sim 7.5$ keV which might be the Fe K edge and an emission line feature at $\sim6.4 $ keV. We applied the absorption based models as well as the relativistic reflection model to fit the excess observed at $\sim 6-8\kev$ band.  \\

\noindent (1) {\textit{Absorption based models}} \\

In order to include these features, we used an absorption {\tt edge} component ({\tt edge} in {\tt XSPEC}) which improved the fit by $\Delta \chi^2=-41$ for the continuum model 3A and $\Delta \chi^2=-28$ for the continuum model 3B for 2 additional parameters.  Examination of the fit residuals suggested another absorption feature at $\sim7\kev$. By the application of a second absorption {\tt edge} the fit was improved by $\Delta\chi^2$=$-28$ for the continuum model 3A and $\Delta\chi^2$=$-22$ for the continuum model 3B along with the first {\tt edge} (for 2 additional parameters). The two edges at $\sim7.65$ and $\sim6.92\kev$ were comparable in their depths. The best-fitting parameters of the new models 3A(+2 edges) and 3B(+2 edges) (see Table~\ref{models} for description) are listed in Table~\ref{suzaku_fit_pars}. The spectral data and deviations of the observed data from the best-fitting models with the $\chi$ value multiplied with a single and double edges are shown in Fig.~\ref{fig_suzaku_fit}(b) and (c). \\

Two absorption {\tt edge} models were also employed with the continuum model 2. The new model reads as model 2(+2 edges), see Table~\ref{models} for description. Here also two absorption edges with energies $7.7\pm0.13\kev$ and $6.9\pm0.07\kev$ provided us a satisfactory fit ($\chi^2/dof = 406/327$). The best-fitting cutoff power-law $\Gamma$ and high energy cutoff values were found to be $1.0\pm0.4$ and $6.1_{-1.1}^{+2.5}\kev$, respectively. Next we added two {\tt edge} models with the continuum model 4 and the new model 4(+2 edges) (see Table~\ref{models} for description) resulted in $\chi^2/dof = 405/326$ with the best-fitting seed photon temperature $kT_{seed}=0.91_{-0.10}^{+0.06}\kev$, electron temperature $kT_{e}=3.4_{-0.15}^{+0.27}\kev$ and optical depth $\tau\gtsim11.7$. Here also edge parameters were similar to those found in the case of other continuum models. All the best-fitting parameter values of the model 2(+2 edges) and model 4(+2 edges) are reported in Table~\ref{fit_continuum}.
We also applied two {\tt edge} component to the continuum model 0A and this model 0A(+2 edges) (see Table~\ref{models}), provided  $\chi^2$/dof=$436.6/328$. Thus, model 3A(+2 edges) or 2(+2 edges) or 4(+2 edges) provided significantly better fit ($\chi^2/dof=405/326$, $406/327$ and $405/326$, respectively) than the model 0A(+2 edges) ($\chi^2/dof = 436.6/328$). Since the {\tt bbody} model component is included in all our best-fitting models, we estimated blackbody area fraction ($A_{bb}$) using the best-fitting blackbody normalization.  $A_{bb}$ is defined as the the fraction of the NS surface area that is emitting as a blackbody (assuming a 10 km radius for the NS.)\\

\begin{figure*}
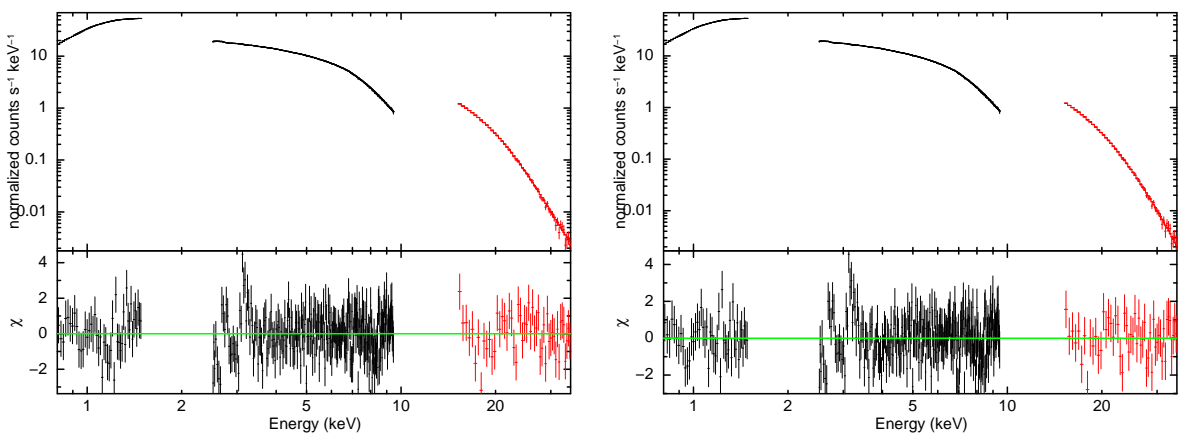

\centering
\includegraphics[scale=0.32,angle=-90]{fig4.ps}
\includegraphics[scale=0.32,angle=-90]{fig5.ps}
\caption{ Left: \suzaku{} spectrum is fitted with a reflection model {\tt (bbrefl\_Hestar.fits)} along with the continuum model 1, {\tt TBabs$\times$(diskbb + bbody + powerlaw)}. Deviation with respect to the best-fitting model in units of $\sigma$ is shown in the lower panel. Right: photoionized absorption model {\tt zxipcf} is applied to the continuum model 3A to fit the \suzaku{} spectrum, {\tt TBabs$\times$zxipcf(diskbb + bbody + nthcomp)}. Lower panel shows the deviation with respect to the best-fitting model. In both the spectral fitting an excess $\sim 3.2 \kev$ is observed, which resulted large $\chi^{2}/dof$.}
  \label{fig_reflection}
\end{figure*}

Since the models involving two absorption edges provided a good description of the data, we employed a physical photoionized absorption model. We used the {\tt zxipcf} model in {\tt XSPEC} \citep{2008MNRAS.385L.108R,2007A&A...463..131M} after assuming that it covered some fraction of the X-ray source.  We replaced the {\tt edge} components of the model 3A(+2 edges) and 3B(+2 edges) with a single {\tt zxipcf} model. According to this model a power-law source illuminates the photoionized gas of the turbulent velocity of $200$ km s$^{-1}$ and reproduces absorption. The addition of this model to the data improved the fit by $\Delta \chi^2=-54$ and $-43$ for 3 additional parameters, compared to the fit with the model 3A and 3B, respectively (corresponding $\chi^{2}/dof$ values are $419/327$ and $437/327$, respectively). The addition of the same model led to an improvement of $\Delta \chi^2=-55$  for 3 additional parameters (corresponding $\chi^{2}/dof$=$417/327$) compared to the fit with the model 4. In particular, the fit was improved at the energy ranges $6 - 9\kev$ and broad residuals were no longer present in the above mentioned energy range (see Fig.~\ref{fig_reflection}, right panel). We found the best-fitting column density of the photoionized plasma lying in the range $N_{H}\sim(0.6-1.8)\times 10^{23}$ cm$^{-2}$.  The best-fitting ionization parameter (log$\xi$) and the covering fraction lay in the range $0.32 - 0.90$ and $0.21 - 0.49$, respectively. During the fitting with the {\tt zxipcf} model the redshift parameter was frozen to zero. \\
 
However, we noted that  our best-fitting reduced $\chi^{2}$ values were still high because of the residuals present at $\sim3\kev$. In both the cases, the model fits were significantly improved by the addition of a narrow Gaussian at $\sim3\kev$. Thus the new $\chi^2/dof$ values for the models 3A(+2 edges) and 3B(+2 edges) became $358/323$ and $383/323$, respectively. The $2\sigma$ upper limit of the Gaussian width ($\sigma$) was $<0.088$ with the best-fitting line energy $3.2\pm0.03\kev$. \\

\noindent (2) {\textit{Relativistic reflection models}} \\

We also examined if the features in the iron K band were better described by relativistic broad iron line than the edges. For this, we removed the two {\tt edge} components from the model 3A(+2 edges) and added a {\tt LAOR} line component ({\tt LAOR} in {\tt XSPEC}, \citealt{1991ApJ...376...90L}). The fit resulted in $\chi^2/dof=426.6/325$ with best-fitting line parameters $E_{LAOR} = 6.93_{-0.06}^{+0.34}\kev$, inner radius $R_{in} = 4.2_{-0.6}^{+0.8}R_g$, emissivity index $\beta=3.2_{-0.2}^{+0.4}$, equivalent width (EW) $\sim 45{\rm~eV}$. The inclination was not well constrained ($>14$ deg.). \\

The irradiation of hard ionizing flux onto the accretion disc leads to the line emission. In order to model the broad Fe K$\alpha$ line along with reflection continuum, it is necessary to employ a reflection model. Following \citet{2010ApJ...720..205C}, here we employed a reflection model where the blackbody component provides the illuminating flux instead of a power-law spectrum. We used the reflection model {\tt (bbrefl\_Hestar.fits)}\footnote{https://heasarc.gsfc.nasa.gov/xanadu/xspec/models/bbrefl.html} appropriate for a $He$-star donor, provided by \citet{2004ApJ...602L.105B} where a constant density slab is illuminated by a blackbody. The reflection model was relativistically blurred by {\tt rdblur} model in XSPEC. We applied this reflection model with the continuum model 1 and this model, {\tt TBabs$\times$(bbody + diskbb + rdblur$\times$bbrefl\_Hestar.fits + powerlaw)} provided us a good fit ($\chi^2/dof = 429/326$) with the best-fitting parameters: inner radius $R_{in} = 8.3_{-1.5}^{+1.0}R_g$, inclination $i=27\pm2$ deg., ionization parameter log$\xi=3.1\pm0.4$, emissivity index $\beta>4.7$. These parameters are consistent within errors with that obtained by \citet{2010ApJ...720..205C}. The spectral data and residuals (in units of $\sigma$) are shown in Fig.~\ref{fig_reflection} (left panel). However, the reported $\chi^2/dof$ value was high because of the residuals at $\sim3.2\kev$. If we removed the energy band $2.5-3.5\kev$ from the spectral fitting, the same model provided $\chi^2/dof = 316/290$ without changing the spectral parameters significantly. \\

We employed the same reflection model convolved with {\tt rdblur} to the continuum model 2 which resulted in $\chi^2/dof = 420/325$ with the best-fitting parameters: inner radius $R_{in} = 8.5_{-1.6}^{+2.5}R_g$, inclination $i=26.4_{-3.0}^{+2.6}$ deg, ionization parameter log$\xi=2.6\pm0.2$.
So far the reflection model was not applied with the continuum models which contains thermal Comptonization model ({\tt nthcomp} or {\tt CompTT}). Therefore, the same reflection model with {\tt rdblur} was applied along with the continuum model 4 also. It resulted in $\chi^2/dof = 435/325$ with the best-fitting parameters: inner radius $R_{in} = 7.5_{-1.4}^{+0.6}R_g$, inclination $i=28.3_{-3.7}^{+2.6}$ deg, ionization parameter log$\xi=3.1\pm0.3$ .\\

\begin{table*}
 \caption{Spectral models used.} 
\begin{tabular}{|p{3.5cm}|p{10.0cm}|}
\hline
Model Name & Model Components \\
\hline
Model 0A(+2 edges) & {\tt TBabs$\times$edge$\times$edge(diskbb $+$ nthcomp)} \\[-0.25cm]
Model 2(+2 edges) & {\tt TBabs$\times$edge$\times$edge(diskbb $+$ bbody $+$ cutoffpl)} \\[-0.25cm]
Model 3A & {\tt TBabs$\times$(diskbb $+$ blackbody $+$ nthcomp)} with $kT_{seed}=kT_{disc}$ \\[-0.25cm]
Model 3B & {\tt TBabs$\times$(diskbb $+$ blackbody $+$ nthcomp)} with $kT_{seed}=kT_{bb}$ \\[-0.25cm]
Model 3A(+2 edges) & {\tt TBabs$\times$edge$\times$edge(diskbb $+$ blackbody $+$ nthcomp)} with $kT_{seed}=kT_{disc}$ \\[-0.25cm]
Model 3B(+2 edges) & {\tt TBabs$\times$edge$\times$edge(diskbb $+$ blackbody $+$ nthcomp)} with $kT_{seed}=kT_{bb}$ \\[-0.25cm]
Model 4(+2 edges) & {\tt TBabs$ \times$edge$\times$edge(diskbb $+$ bbody $+$ CompTT)} with $kT_{seed}=kT_{disc}$ \\
\hline
\end{tabular}\label{models} \\
\end{table*}

Finally, in the Table~\ref{chi2}, we enlisted $\chi^{2}(dof)$ values that we obtained after addition of line model or absorption based model or reflection model with different continuum models in the \suzaku{} spectra of 4U~1820--30. However, we found that the emission/absorption features observed $\sim 6-8\kev$ is moderately better described by two absorption edge models compared to the blurred reflection model or partially covering absorption model.
The weak absorption features seen at $\sim1\kev$ and $\sim 2.5\kev$ and $\sim3\kev$ (see Fig.~\ref{fig_suzaku_fit}) in the \suzaku{} spectrum were unidentified features and the origin of these features may be instrumental. We did not try to model these weak absorption features as any support for the absorption based model from these features is unexpected.

\subsection{\nustar{} \& \swift{} spectra}

We simultaneously fitted spectral data of \nustar{} FPMA and FPMB in the $3-79$ keV energy band and \swift{}/XRT spectrum in the $0.3-8$ keV energy band. To fit both the soft and hard components in the spectra, we selected a model consisting of the blackbody ({\tt bbody}) and the power-law model ({\tt powerlaw}). As before, we used the {\tt TBabs} model to account for the Galactic absorption.  A constant was used to account for relative normalization between different instruments. This model failed to describe the spectrum with $\chi^2$/dof of $856/702$. To see whether multiple soft components were required, we added the disc blackbody component ({\tt diskbb}). The overall fit improved significantly and this model (model 1) resulted in $\chi^2$/dof = $742/700$. To check whether a cut-off exists in the spectrum, we replaced the power-law model with a cutoff power-law ({\tt cutoffpl}). Although the model (model 2) provided  an acceptable fit ($\chi^2/dof = 711/699$), the power-law photon index became negative and the cutoff energy was below 5 keV, which were unacceptable. \\

\begin{figure}
\centering
\includegraphics[scale=0.35,angle=-90]{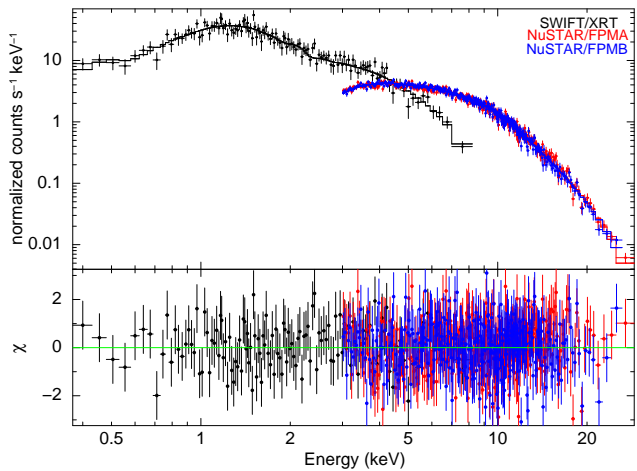}
\caption{Joint spectral fitting of \nustar{} FPMA and FPMB and \swift{} XRT. The spectral data set with the
    best-fitting spectral model {\tt TBabs$\times$(diskbb + bbody + nthcomp)}, deviations of the observed data from the best-fitting model (lower panel).}
  \label{fig_nustar_fit}
\end{figure}

Low reduced $\chi^2$ with cut-off power-law is most likely a signature of the presence of thermal Comptonization. Therefore, we replaced the cut-off power-law model by the thermal Comptonization model ({\tt nthcomp}). The combination of this thermal Comptonization model and blackbody (model 0B) provided an exceptionally good fit ($\chi^2/dof$=$716/701$), with the seed photon temperature tied to the blackbody $kT_{bb}\sim 0.53\pm0.04\kev$ and $\Gamma\sim1.8\pm0.03$. When we replaced the blackbody model with the multi-color disc blackbody model, the fit still held equally good along with {\tt nthcomp} model (model 0A). However, we retained both the blackbody and the multi-temperature disc blackbody models for two reasons : (1) \suzaku{} spectral fitting requires disc component with high significance. (2) The origin of multi-color disc blackbody component is due to photons from accretion disc while the origin of the blackbody component is at the surface of the neutron star. In many cases, both of these are required to fit the spectrum. 
The temperature of seed photons was tied either to multi-temperature disc (model 3A) or to surface blackbody temperature (model 3B). Both of these provided an equally good fit while differences in fluxes were observed. As done before for \suzaku{} spectrum, we tried to fit the spectrum with another thermal Comptonization model {\tt CompTT} by \citet{1994ApJ...434..570T} along with two thermal components (model 4). This fit resulted in similar electron temperature ($kT_{e}$) and the seed photon temperature ($kT_{seed}$) as derived by the {\tt nthcomp} model with an optical depth ($\tau$) of the Comptonizing cloud $\sim9.5$ (see Table~\ref{fit_continuum}).  
 Table~\ref{suzaku_fit_pars} lists the best-fitting parameters of the best-fitting model 0A, 0B, 3A and 3B for the joint \nustar{} and \swift{} spectra and Fig.~\ref{fig_nustar_fit} shows the simultaneous FPMA, FPMB and XRT spectral fitting, the best-fitting model and $\chi$ residuals. Addition of an absorption {\tt edge} or a {\tt Gaussian} or a {\tt diskline} did not improve the fit. 
However, we introduced two edges at energies which were within the ranges consistent with that obtained from the \suzaku{} spectral fitting keeping optical depth ($\tau$) as a free parameter. In both the models 3A and 3B, for the edge in the energy range $7.54 - 7.80$ keV, $90\%$ upper limit on the optical depth ($\tau$) was $ 0.094$ which was within errors of the depth measured with \suzaku{}. For the second edge in the energy range $6.86 - 6.98$ keV, the $90\%$ upper limit on the optical depth ($\tau$) was $<0.042$. The upper limit of $\tau$ in the fits appeared to be consistent with each other.

\begin{table*}
  \caption{ Energy spectral parameters of \suzaku{} (combined spectra of XIS ($0.8-10$ keV) and HXD/PIN ($15-35$ 
    keV)) and \nustar{} + \swift{} (combined spectra of \nustar{} and \swift{}/XRT), fitted with different model combinations.
    Quoted errors are at $90\%$ confidence level.} 
  
  \begin{tabular}{lllllllll}
    \hline
    Component     & Parameter & \multicolumn{3}{c}{\it Suzaku} & \multicolumn{4}{c}{\it NuSTAR + Swift/XRT} \\  
    \cline{3-9} 
    &           &  Model 0A  & Model 3A & Model 3B & Model 0A & Model 0B & Model 3A & Model 3B \\[-0.3cm]
    &           & (+2 edges) & (+2 edges) & (+2 edges) &    &     &    &     \\
    \hline
    {\scshape tbabs}    & $N_{H}$($\times 10^{22} cm^{-2}$) &$0.19\pm0.004$ & $0.18\pm0.004$ & $0.21\pm0.03$ & $0.15$ (f) &  $0.15$ (f)  &  $0.16$ (f) & $0.15$ (f) \\
    {\scshape edge(1)}  & $E_{edge}$ & $7.67\pm0.14$ & $7.67\pm0.13$  & $7.66\pm0.14$ & -- & --  & -- & -- \\
    & $\tau_{max}$ ($\times 10^{-2}$) &$3.4\pm1.1$   &  $3.6\pm1.0$ &  $3.2_{-1.1}^{+0.6}$ & --  & --  & -- & -- \\
    {\scshape edge(2)}  & $E_{edge}$ &$6.93\pm0.07$  & $6.92\pm0.06$   &  $6.92\pm0.07 $ & -- & -- & -- & --          \\
    & $\tau_{max}$ ($\times 10^{-2}$) & $3.3\pm1.0$  &  $3.6\pm0.9$ & $3.4_{-0.5}^{+0.9}$ & -- & -- & --  & -- \\
    {\scshape diskbb}   & kT$_{disc}$(keV) & $0.93\pm0.03$ &   $0.98\pm0.03$ & $0.44_{-0.06}^{+0.10}$ & $0.66_{-0.07}^{+0.09}$  &   --  &  $0.85_{-0.17}^{+0.09}$  & $0.84\pm0.08$ \\
    & $N_{DBB}$~$^a$   &$142\pm5$   & $115_{-109}^{+45}$  &  $1951_{-942}^{+1421}$ & 105.5$^{+50.5}_{-36.9}$ & --   & $160_{-39}^{+67}$ & $183_{-65}^{+72}$         \\
    {\scshape blackbody} & $kT_{bb}$($\kev$) & --  &   $2.46_{-0.08}^{+0.04}$ & $0.65\pm0.05$ & --  & 0.53$\pm0.04$  &$1.64_{-0.28}^{+0.23}$ & $1.59_{-0.05}^{+0.07}$ \\
    & $N_{BB}$~$^b$($\times 10^{-2}$)   & --  &   $6.04_{-1.40}^{+0.91}$ & $1.35_{-0.17}^{+0.19}$ & --   & $0.25\pm0.01$  &   $1.50_{-0.50}^{+2.30}$ & $1.10_{-0.40}^{+1.90}$ \\

    {\scshape nthcomp} & $\Gamma$ & $1.43\pm0.03$ & $2.16_{-0.93}^{+1.50}$ & $1.47\pm0.04$ & $1.78\pm0.03$ & $1.77\pm0.03$ & $1.58_{-0.05}^{+0.22}$  & $1.97_{-0.25}^{+0.15}$  \\
    & $kT_{e}$ (keV) &$2.80\pm0.02$  & $\ge 2.08$  &  $2.82\pm0.02$ &$2.95\pm0.04$ & $2.93_{-0.04}^{+0.02}$ & $3.04_{-0.22}^{+0.15}$ & $2.98_{-0.19}^{+0.25}$ \\
    & $kT_{seed}$ (keV) &$=kT_{disc}$  & $=kT_{disc}$ & $=kT_{bb}$ & $=kT_{disc}$  & $=kT_{bb}$ & $=kT_{disc}$  & $=kT_{bb}$    \\
    & inp type &$1.0$(f)  &  $1.0$ (f) &  $0.0$ (f)  & $1.0$ (f)   & $0.0$(f)  & $1.0$ (f) & $0.0$(f)        \\
    & $n_{nthcomp}$~$^c$  & $0.38\pm0.04$ &  $0.37\pm0.11$ & $0.22\pm0.05$  & $0.42_{-0.11}^{+0.33}$ & $0.48_{-0.05}^{+0.03}$ & $0.11\pm0.05$  & $0.05\pm0.02$  \\ \hline
    diskbb  & $R_{in}{\rm~(km)}$ & $10.3\pm0.2$ & $9.3^{+1.9}_{-6.1}$ & $38.3^{+12.0}_{-10.8}$ & $8.9^{+1.9}_{-1.7}$ &  -- & $11.0^{+2.1}_{-1.5}$ & $11.7^{+2.1}_{-2.3}$ \\
    & $f_{diskbb}$($\times10^{-9}$ cgs) & $2.3\pm0.01$  & $2.4\pm0.01$   & $1.6\pm0.01$  & $0.4\pm0.05$  & -- & $1.8\pm0.07$ & $2.1\pm0.05$ \\
    & $f_{bb}$($\times10^{-9}$ cgs)    & --  &  $5.1\pm0.01$   &  $1.2\pm0.01$ &  --  &  $0.21\pm0.03$  & $1.3\pm0.08$ & $1.6\pm0.07$ \\
    BB area fraction~$^d$ & $A_{bb}$         &  --  & $0.08\pm0.01$    &  $3.68\pm0.98$   &  -- & $1.45\pm0.37$  & $0.10\pm0.04$  & $0.14 \pm0.02$  \\ 
    & $f_{seed}$($\times10^{-9}$ cgs) &  $2.2\pm0.01$   & $1.9\pm0.01$  & $3.4\pm0.01$ & $1.7\pm0.05$ & $2.2\pm0.03$ & $0.5\pm0.07$ & $0.9\pm0.05$  \\
    & $f_{nthcomp}$($\times10^{-9}$ cgs) & $9.1\pm0.02$  & $4.2\pm0.01$  & $8.8\pm0.02$ & $4.9\pm0.02$ & $5.5\pm0.01$ & $2.7\pm0.01$ & $2.4\pm0.01$ \\
Amp. fac.~$^e$ &  A & $4.1\pm0.01$ & $2.2\pm0.01$ & $2.6\pm0.002$  &  $2.9\pm0.07$ & $2.5\pm0.03$ & $4.5\pm0.74$ & $2.6\pm0.14$ \\
  
     \hline 
    & $\chi^{2}/dof$ &$436.6/328$  & $405/326$  & $430/326$  & $717/701$  & $716/701$ & $712/699$ & $711/699$ \\
    \hline
  \end{tabular}\label{suzaku_fit_pars} \\
  Spectral models - Model 0A: {\tt TBabs$\times$(diskbb $+$ nthcomp)};  
    Model 0B: {\tt TBabs$\times$(bbody $+$ nthcomp)};  
    Model 3A: {\tt TBabs$\times$(diskbb $+$ bbody $+$ nthcomp)} with $kT_{seed}=kT_{disc}$;  
    Model 3B: {\tt TBabs$\times$(diskbb $+$ blackbody $+$ nthcomp)} with $kT_{seed}=kT_{bb}$ \\
    Model 3A/3B(+2 edges): {\tt TBabs$\times$edge$\times$edge(diskbb $+$ blackbody $+$ nthcomp)} with $kT_{seed}=kT_{disc}$/$kT_{seed}=kT_{bb}$ \\
   $^{a,}$$^{b,}$$^{c}$ Normalization parameter of the {\tt diskbb}, {\tt blackbody} and {\tt nthcomp} component, respectively. \\
   $^d$ BB area fraction is the fraction of the NS surface area that is emitting as a blackbody (assuming a $10$ km radius of the NS.)\\  $^e$ Amplification factor which is the ratio of the Comptonized luminosity to the seed photon luminosity.\\
 \end{table*}

\begin{table*}
  \caption{Best-fitting spectral parameters for different applied models in the \suzaku{} and \nustar{} + \swift{} spectrum of 4U~1820--30} 
  
    \begin{tabular}{|p{2.0cm}|p{2.2cm}|p{1.8cm}|p{1.8cm}|p{1.8cm}|}
    \hline
    Component & Parameter & \multicolumn{2}{c}{\it Suzaku} & \multicolumn{1}{c}{\it NuSTAR+Swift/XRT}\\
    \cline{3-5} 
              &            & Model 2 & Model 4 & Model 4 \\[-0.3cm] 
              &            & (+2 edges) & (+2 edges) &       \\   
    \hline
    {\scshape tbabs} & $N_{H}(\times 10^{22}cm^{-2})$ & $0.21\pm0.02$ &$0.19\pm0.01$  & $0.16$(f) \\
    {\scshape diskbb} & kT$_{disc}$ (\kev) & $1.04\pm0.09$ & $0.91_{-0.10}^{+0.06}$ & $0.60_{-0.11}^{+0.30}$ \\
     & $N_{DBB}$   & $114_{-22}^{+40}$  & $254_{-44}^{+97}$ & $593_{-433}^{+373}$\\
    {\scshape blackbody} &  kT$_{bb}$ (\kev) & $2.49\pm0.05$ &$2.30\pm0.06$  & $1.90_{-0.49}^{+0.26}$\\
     & $N_{BB}$ ($\times 10^{-2}$)& $6.04\pm0.9$ & $5.18_{-0.97}^{+0.74}$ & $0.70\pm0.6$\\
    {\scshape cutoffpl} & $\Gamma$ & $1.0\pm0.4$ & -- & --\\
     & HighEcut & $6.1_{-1.1}^{+2.5}$ & -- & --\\
     & norm  & $0.39\pm0.09$  & -- & --\\
    {\scshape comptt} & $kT_{seed}$(\kev) & -- & $=kT_{disc}$ & $=kT_{disc}$ \\
     & $kT_{e}$(\kev) & -- & $3.4_{-0.15}^{+0.27}$ & $3.2_{-0.29}^{+0.61}$ \\
     & $\tau$ & -- & $\gtsim11.7$ & $\gtsim9.5$ \\
     & norm ($\times 10^{-26}$) & -- & $0.14_{-0.05}^{+0.08}$ & $0.20_{-0.15}^{+0.08}$ \\
    \hline
     & $\chi^{2}/dof$ & $406/327$ & $405/326$ & $710/700$\\
    \hline 
  \end{tabular}\label{fit_continuum} \\
  {\footnotesize Model 2(+2 edges): {\tt TBabs$\times$edge$\times$edge(diskbb $+$ bbody $+$ cutoffpl)};
  Model 4: {\tt TBabs$\times$(diskbb $+$ bbody $+$ CompTT)} with $kT_{seed}=kT_{disc}$; Model 4(+2 edges): {\tt TBabs$\times$edge$\times$edge(diskbb $+$ bbody $+$ CompTT)} with $kT_{seed}=kT_{disc}$\\
 {\bf Note}: For both the models in the \suzaku{} spectral fitting, two absorption edges are well fitted at energies $6.92\pm0.07\kev$ and $7.68\pm0.14\kev$. If we introduce a Gaussian line at $\sim3.2\kev$ then $\chi^{2}/dof$ values becomes $356/324$ and $359/324$ for the model 2(+2 edges) and model 4(+2 edges), respectively . In {\tt CompTT} model we considered only the spherical geometry.}
 \end{table*}

\begin{table*}
  \caption{ List of $\chi^{2}(dof)$ values on addition of line model or absorption model or reflection model with the different continuum models to the \suzaku{} spectrum of 4U~1820--30.} 
  
   \begin{tabular}{|p{6.5cm}|p{1.2cm}|p{1.2cm}|p{1.2cm}|p{1.2cm}|}
    \hline
                &      \multicolumn{4}{c}{$\chi^{2}(dof)$}  \\ 
     \cline{2-5}
    Continuum models   &  LAOR  & Blurred reflection & zxipcf & 2 edges \\
    \hline
    Model 1: {\tt TBabs$\times$(diskbb $+$ bbody $+$ powerlaw)} & $431.3(326)$ & $429(326)$ & $427.7(328)$ & $416.4(327)$   \\
    Model 2: {\tt TBabs$\times$(diskbb $+$ bbody $+$ cutoffpl)} & $423.7(326)$ & $420(325)$ & $422.7(328)$ & $406(327)$   \\
    Model 3$^{*}$: {\tt TBabs$\times$(diskbb $+$ bbody $+$ nthcomp)} & $426.6(325)$ & $431.5(325)$ & $419(327)$ & $405(326)$   \\
    Model 4: {\tt TBabs$\times$(diskbb $+$ bbody $+$ CompTT)} & $429.2(325)$ & $435(325)$ & $417(327)$ & $405(326)$   \\
    \hline
\end{tabular}\label{chi2} \\
  {\footnotesize ~$^{*}$ Note that LAOR, Blurred reflection, zxipcf and 2 edges were applied to the model 3 when $kT_{seed}=kT_{disc}$ (model 3A).} \\
    \end{table*}

\section{Discussion}
We examined the continuum behaviour as well as the presence of the broad iron line in the X-ray spectra of 4U~1820--30 using broadband observations with \suzaku{}, \nustar{} and \swift{}. We used a number of continuum models to the data and investigated the presence of discrete features in the iron K band. Studies of Fe fluorescence in X-ray binaries are not trivial and issues regarding statistics, bandpass, and competing physical processes play vital roles in the interpretation of the results. Our spectral components included the emission from the NS surface and an accretion disc around the NS, thermal Comptonization in a corona of hot electrons, reflection off the accretion disc and photoionized absorption from material in the vicinity of the system.
From our spectral analysis we did not observe any trace of Fe K$\alpha$ emission line in the energy range $5.8-7.0\kev$ using simultaneous spectroscopy of \nustar{} and \swift{}. But at the same time, the \suzaku{} XIS data clearly showed absorption or emission features in the iron K band (see Fig.~\ref{fig_suzaku_fit} and Fig.~\ref{fig_edge}). These features can be described in two ways -- ($i$) absorption edges or absorption from photoionised gas, and ($ii$) blurred reflection. The absorption features observed in the data  at $\sim 6.9\kev$ and $\sim 7.6 \kev$, when fitted with two {\tt edge} components, yielded a better fit statistically.
It may be noted that, \citet{2010ApJ...720..205C} detected a significant broad iron line in the \suzaku{} observations of the source 4U~1820--30 but the combined \xmm{} and \inte{} spectrum did not find clear evidence of iron emission lines (see \citealt{2012A&A...539A..32C}). Using \rxte{} data of the source 4U~1820--30, \citet{2013ApJ...767..160T} reported a broad Fe emission line during both the soft and the hard states, although \rxte{} is much less sensitive in detecting emission lines compared to high resolution instruments like \nustar{} or \suzaku{}. \\

It is important to note that same edge energies ($E_{edge}$) were recovered from the different continuum models. Thus, the choice of continuum did not affect the values of the absorption edge related parameters. We discuss here alternative interpretations of the broad feature seen $\sim 6.4\kev$ in the \suzaku{} spectrum of 4U~1820--30. A broad emission line was no longer required if we introduced two absorption edges at energies above $6.9\kev$.  \citet{2005astro.ph.11072D} applied two absorption edge models, instead of a broad emission line, to fit broad line feature observed $\sim6.5\kev$ in the {\it chandra} and {\it RXTE} observation of another atoll type low mass NS X-ray binary 4U~1728--34.  
 The broad features seen in the \suzaku{} spectra at $\sim 6 - 9 \kev$ was also well described by a blurred reflection model. From the reflection model fit, the inner disc radius and the source inclination were found to be $\sim8\ GM/c^{2}$ and $\sim28$ deg., respectively. Thus, it is clear that the presence of absorption edges may mimic the broad iron line like feature or vice-versa (see also \citealt{2011A&A...530A..99E}). \\
 
It may also be noted that the Fe K-edge is observed in the $7.1-9.1$ keV range, and our observed energy for one edge at $6.92\pm0.07\kev$ was marginally lower than the K-edge energy of neutral iron. Thus, the edge must arise close to the surface of the neutron star which might then be gravitationally redshifted to observed energies. The required redshifts for $\sim6.9\kev$  and $\sim7.7\kev$ edges are $0.025$ and $0.20$, respectively. For the aforesaid redshifts, the absorbing material must be located at a distance $3.4-20.7\: R_S$ from the neutron star. This suggests that the absorption edge must arise from close to the neutron star. Since there was no evidence for X-ray bursts during the \suzaku{} observation, the absorption features were unlikely to be the burst driven wind. On the other hand, since the absorption features were described by photoionised {\tt zxipcf} model ($N_{H}\sim(0.6-1.8)\times 10^{23} {\rm~cm^{-2}}$ and log $\xi\sim 0.32-0.90$), the absorption features might arise due to the winds from the inner accretion disc \citep{2005A&A...436..195B}. The source was observed in a high/soft state during the \suzaku{} observation and the luminosity of the source lies $\sim(7-8)\times10^{37}{\rm~ergs\ s^{-1}}$ (as bright as persistent atolls, e.g. GX 13+1, GX 9+1). The accretion rate onto the compact object is very high $\sim1.1\times10^{18}{\rm~gm\ s^{-1}}$ (\citealt{2013ApJ...768..184H}) which is comparable to the bright atoll GX 13+1 ($\sim10^{18}{\rm~gm\ s^{-1}}$, \citealt{2004ApJ...609..325U}) and ultracompact binary 4U~1916--053 ($\sim0.6\times10^{17}{\rm~gm\ s^{-1}}$, \citealt{2013ApJ...768..184H}) for whom disc winds were detected (\citealt{2012A&A...543A..50D, 2006A&A...445..179D}). This suggests that the high accretion rate may give rise to massive, radiatively-driven winds. However, it may be noted that the absorption by a photo-ionized disc wind was also observed in sources that accreates matter lower than those of GX 13+1 and GX 9+1 \citep{2005A&A...436..195B,2006A&A...445..179D}\\

The \suzaku{} and \swift{}+\nustar{} broadband datasets also allowed us to investigate the continuum models. It is well known that a number of different continuum models describe the observed spectra of neutron star LMXBs \citep{1988ApJ...324..363W,1989PASJ...41...97M, 2001AdSpR..28..307B, 2010ApJ...720..205C}. Our analysis of broadband \suzaku{} data showed that the continuum model consisting of an accretion disc blackbody, a simple blackbody and thermal Comptonization of disc photons best described the observed data statistically (see Table~\ref{suzaku_fit_pars}). The same model also described the \nustar{} data but it was not possible to distinguish statistically the models with diskbb and blackbody seeds for thermal Comptonization (compare with \citealt{2007ApJ...667.1073L}). 

In the Comptonization model, we assumed two different origins of the input seed photons for thermal Comptonization to investigate the emission geometry of the source (as done by \citealt{2013MNRAS.432.1144S,2014MNRAS.440.1165L}). For the models 0A, 0A(+2 edges), 3A and 3A(+2 edges) we assumed that the input photon distribution was due to multicolor disc blackbody {\tt(diskbb)} component ($kT_{seed}=kT_{disc}$) and for the rest of the models we assumed it due to single temperature blackbody {\tt(bbody)} component ($kT_{seed}=kT_{bb}$). We calculated the inner accretion disc radius from the normalization parameter of the disc blackbody ({\tt diskbb}) component. It gave us physically acceptable inner disc radius for all the models ($\gtsim 10~{\rm~km}$). We also calculated the fraction of the neutron star surface area that was emitting as blackbody emission by assuming that the neutron star had a radius of $10{\rm~km}$.  Clearly, in this scenario the blackbody area could not exceed the neutron star surface area, and models 0B and 3B(+2 edges) with fractional BB area of $1.45$ and $3.68$, respectively, were ruled out. Thus, it is unlikely that the blackbody is providing the seed photons for the thermal Comptonization.  \\
  
In all the models where we set $kT_{seed}=kT_{disc}$, the seed photon flux was consistently less than the diskbb flux ($f_{seed}/f_{diskbb} < 1$). It was a good consistency check as in all the models we assumed that the input photon arises from the accretion disc.  In these models, the blackbody emission is arising from a small fraction ($8\%$ to $14\%$) of the neutron star surface. This was similar to the result obtained by \citet{2007ApJ...667.1073L} based on \rxte{} observations of Aql~X--1 and 4U~1608--52. These results indicated that a standard thin accretion disc impacted the neutron star where the colliding gas formed the boundary layer covering a substantial area of the neutron star surface and radiated efficiently via thermal Comptonization. In this scenario, the blackbody emission might arise from the uncovered polar regions as the energy deposited by the impact of the disc would quickly get distributed due to thermal conduction. 
  This scenario was supported by the  small inner disc radii $R_{in} \sim 10{\rm~km}$ inferred from the models 0A(+2 edges) and 3A(+2 edges).  In a geometry in which the disc provided the seed photons for thermal Comptonization in a boundary layer between the innermost disc and the surface of the neutron star, the ratio of seed flux to the disc flux could not exceed unity and it was expected to be substantially less than one if the boundary layer was confined in a narrow ring in equatorial plane. For the model 3A, we found $f_{seed}/f_{diskbb} \sim 0.79$ (\suzaku{}) and $\sim 0.28$ (\swift{} + \nustar{}), it was possible that the Comptonizing corona might be covering substantial part of the neutron star surface as suggested by the blackbody emission from the polar regions or additionally the corona might be covering the innermost part of the disc which in turn resulted in highly Comptonized spectrum as observed.  \\

The size and geometry of the boundary layer is an open issue which is not satisfactorily understood (see e.g.,\citealt {1986SvAL...12..117S, 1999AstL...25..269I, 2001ApJ...547..355P, 2002AstL...28..150G}). Numerical calculations in the classical theory of boundary layers predict that boundary layer can inflate in a radial direction in order to radiate away excess luminosity during high accretion rate state. However, at low accretion rate which is usually observed during hard state, inflation because of radiation pressure is negligible and the boundary layer can be assumed to be compact within the size of the fraction of the neutron star radius \citep{2001ApJ...547..355P}. However, the amount of seed photon flux and the Comptonized flux obtained in this work is not consistent with a narrow and thin boundary layer region. Additionally, for a face-on system, boundary layer emission is supposed to be very strong during hard state but the present system is more edge on. \\

In an alternative and more realistic formalism by \citet{1999AstL...25..269I}, the boundary layer is found to spread onto the surface of the neutron star covering a fraction depending on the mass accretion rate. Unlike the standard model where the accretion energy is concentrated in a small radius, the \citet{1999AstL...25..269I} model allows the dissipation of energy in a sufficiently broad region of NS surface. This is in fair agreement with the high Comptonized flux as well as seed photon flux as reported in the present paper. Broadband X-ray observations such as those would be possible with the upcoming {\it Astrosat} mission will play an important role in further understanding the geometry of the X-ray emitting regions in accreting neutron star LMXB. \\

\section{Acknowledgements}
We are thankful to the anonymous referee whose detail indepth comments helped us in improving the paper. 
This research has made use of data and/or software provided by the High Energy Astrophysics Science Archive Research Centre (HEASARC). Aditya S. Mondal would like to thank Inter-University Centre for Astronomy and Astrophysics (IUCAA) for hosting him during subsequent visits. BR likes to thank IUCAA for hospitality and other facilities extended to him during his visit under their visiting Associateship program.

\def\apj{ApJ}
\def\apjl{ApJl}
\def\pasp{PASP} \def\mnras{MNRAS} \def\aap{A\&A} \def\physerp{PhR} \def\apjs{ApJS} \def\pasa{PASA}
\def\pasj{PASJ} \def\nat{Nature} \def\memsai{MmSAI} \def\araa{ARAA}
\bibliographystyle{mn2e}
\bibliography{aditya}

\end{document}